\documentclass{aa}
\usepackage[varg]{txfonts}
\usepackage{graphicx}
\usepackage{rotating}
\usepackage{natbib}
\usepackage{float}
\usepackage{color}
\usepackage{textcomp,gensymb}
\usepackage{array,amsmath}
\usepackage{mathtools}
\usepackage{multirow,makecell}
\usepackage{enumitem}
\usepackage{longtable}
\usepackage{hyperref}
\providecommand{\nolinenumbers}{}
\providecommand{\modulolinenumbers}[1]{}

\begin{document}
\title{Source identification for the \textit{Swift}-BAT 150--month hard X-ray catalog using soft X-ray observations}

\author{K.~Imam\inst{1} \and
N.~Torres-Alba\inst{2} \and
S.~Marchesi\inst{3,1,4} \and
M.~Ajello\inst{1} \and
S.~Joffre\inst{1} \and
I.~Cox\inst{1} \and
A.~Pizzetti\inst{5,1}\fnmsep\thanks{ESO Fellow} \and
X.~Zhao\inst{6,7} \and
A.~Segreto\inst{8,9} \and
A.~Banerjee\inst{1} \and
I.~Pal\inst{1} \and
V.~E.~Gianolli\inst{1} \and
D.~Stern\inst{10}}

\institute{Department of Physics and Astronomy, Clemson University, Kinard Lab of Physics, Clemson, SC 29634-0978, USA\and
Department of Astronomy, University of Virginia, P.O. Box 400325, Charlottesville, VA 22904, USA\and
Dipartimento di Fisica e Astronomia (DIFA), Università di Bologna, via Gobetti 93/2, I-40129 Bologna, Italy\and
INAF – Osservatorio di Astrofisica e Scienza dello Spazio di Bologna, Via Piero Gobetti 93/3, 40129 Bologna, Italy\and
European Southern Observatory, Alonso de Córdova 3107, Casilla 19, Santiago 19001, Chile\and
Department of Astronomy, University of Illinois at Urbana-Champaign, Urbana, IL 61801, USA\and
Cahill Center for Astrophysics, California Institute of Technology, 1216 East California Boulevard, Pasadena, CA 91125, USA\and
INFN, Sezione di Catania, Catania, Italy\and
Istituto di Astrofisica Spaziale e Fisica Cosmica di Palermo (INAF), Palermo, Italy\and
Jet Propulsion Laboratory, California Institute of Technology, Pasadena, CA 91109, USA}

\abstract{
We present a comprehensive catalog of 251 potential counterparts for 250 unassociated hard X-ray sources detected in the \textit{Swift}-Burst Alert Telescope (BAT) 150-month hard X-ray survey. Over 150 months of observation, BAT has detected 2339 sources in the 15$-$150~keV range.  
Among these, 344 do not have a low-energy counterpart. Our study focuses on the analysis of soft X-ray observations (i.e., at energies $<$10 keV) spatially overlapping with these \textit{Swift}-BAT new hard X-ray sources. Such observations were taken with \textit{Chandra}, \textit{Swift-XRT}, \textit{eROSITA}, and \textit{XMM-Newton}. Within the sample of 251 potential counterparts, 94 (37\%) are identified as AGN and 58 (23\%) as galaxies. The rest of the 99 (40\%) sources include pulsars, cataclysmic variables and unclassified soft X-ray (0.5--10 keV) counterparts. Redshift data is available for 139 out of the 251 sources, and its distribution is in close agreement with the redshift distribution of previous BAT catalogs. We also present the results of a small optical spectroscopy campaign of 9 (out of 58) galaxies. The majority of these turned out to be Seyfert 2 galaxies at a redshift slightly larger than the median of the BAT AGN sample.}


\maketitle
\section{Introduction}\label{sec:intro}
The exploration of the cosmos across various electromagnetic wavelengths has been instrumental in unraveling the mysteries of the Universe, from the dynamics of star formation to the behavior of supermassive black holes (SMBHs) at the centers of galaxies. Among these wavelengths, hard X-rays (15--150 keV) provide a unique window into some of the most extreme and energetic phenomena of the Universe, like the accretion of matter onto supermassive black holes and jet emission from blazars \citep[e.g.,][]{1979ApJ...231L.111L,1984AdSpR...3j.211U, 2022ApJ...940...77M}. Hard X-rays can penetrate dense environments, such as the cores of galaxies, where supermassive black holes reside, or regions where stars are forming. By surveying the sky in hard X-rays, we can peer through obscuring material like dust and gas.

%\citep[e.g.,][]{1979ApJ...231L.111L,1984AdSpR...3j.211U,1989PASJ...41..739W, 2017ApJS..233...17R, 2022ApJ...940...77M, 2023MNRAS.518.2938T}

%\textbf{ Many studies at different wavelengths \citep[e.g.,][]{1985ApJ...297..621A, 2019A&A...623A..79C, 2019ApJ...884..171H} across the years have allowed us to build a scenario where most of this obscuration is due to a distribution of material located in close proximity (1 - 10 pc) of the accreting supermassive black hole, and historically this distribution has been called torus}. 
%The distribution of hard X-ray sources across the sky provides important clues about the evolution of galaxies and large-scale structures in the universe.
%\textbf{ By studying the properties of hard X-ray sources using soft X-ray missions, we can probe the cosmic history of star formation, galaxy mergers, and the growth of supermassive black holes over cosmic time \citep{,,2024ApJ...963...53C}}.

The Burst Alert Telescope  \citep[BAT;][]{2005SSRv..120..143B} on board the Neil Gehrels \textit{Swift} Observatory is primarily designed to detect gamma-ray bursts in the 15--150 keV energy range. While waiting for gamma-ray bursts, it continuously scans the sky and collects data from hard X-ray sources, thus allowing us to study hard X-ray sources  \citep[e.g.,][]{2008ApJ...673...96A,2013ApJS..207...19B, 2018ApJS..235....4O}. As an example, the \textit{Swift}-BAT 70-month catalog has been used to study the correlations between X-ray continuum emission and optical narrow emission lines \citep{2015MNRAS.454.3622B} and the link between active galactic nuclei (AGN) Eddington ratio and narrow-emission-line ratios \citep{2017MNRAS.464.1466O}, among others.

The 105-month BAT catalog \citep{2018ApJS..235....4O} has been used to measure the high energy cutoff for local AGN \citep{2017ApJS..233...17R}, in doing so putting constraints on the physical properties of the hot corona responsible for the X-ray emission \citep{2018Galax...6...44M}. Several other studies \citep[e.g.,][]{2019MNRAS.487.2463W, 2021MNRAS.506.4960H} have been conducted on the properties of the X-ray corona by selecting sources from the 105-month \textit{Swift}-BAT catalog, although a full understanding of the mechanisms behind the coronal emission is still to be achieved. The \textit{Swift}-BAT catalog has also been useful in studying the so-called changing look AGN \citep[e.g.,][]{2023MNRAS.518.2938T, 2023NatAs...7.1282R}, a class of AGN that exhibits dramatic, relatively rapid changes in their observed properties, particularly in their optical or X-ray emission. These changes include transitions between different optical spectral classifications (e.g., from a Type 1 to a Type 2 AGN or vice versa) and significant variations in luminosity.

In the hard X-ray regime (E $>$ 10 keV), the Cosmic X-ray Background \citep[CXB;][]{1962PhRvL...9..439G, 2007A&A...463...79G, ajello2008} is primarily dominated by SMBHs accreting gas at the centers of galaxies. A subset ($\sim10\%$) of these AGN launches relativistic jets, and when these jets align closely with our line of sight, they are identified as blazars. Blazars are divided into two main subclasses: flat-spectrum radio quasars (FSRQs) and BL Lacertae objects (BL Lacs), distinguished by the strength of their optical emission lines, with FSRQs displaying lines stronger than 5{\AA}  in equivalent width, while BL Lacs exhibit weak or absent lines \citep[e.g.,][]{1968Natur.218..663S, 1976ARA&A..14..173S}.
%Hard X-ray surveys are crucial in studying these sources, as their spectral energy distributions (SEDs) at these energies are dominated by jet emission driven by inverse Compton (IC) scattering, where relativistic electrons in the jet upscatter low-energy photons. 
Previous studies \citep[e.g.,][]{2009ApJ...699..603A, 2020ApJ...896..172T} have shown that blazars, especially the FSRQ subclass, increase in numbers with redshift in this energy range, indicating that they were more numerous and/or luminous at earlier cosmic times. A study by \cite{2022ApJ...940...77M} analyzed \textit{Swift}-BAT blazars and their jets across cosmic time, finding that FSRQ-dominated blazars evolve positively up to z $\sim$ 4.3, but are less abundant at high redshifts than previously estimated. Blazars contribute 5\%–18\% to the cosmic X-ray background (in the 14--195\,keV energy range) and could account for nearly 100\% of the MeV background \citep{2009ApJ...699..603A,2022ApJ...940...77M}, making them key contributors to the unresolved MeV gamma-ray population. These findings refined our understanding of blazar evolution, jet physics, and their contribution to cosmic backgrounds.

The \textit{Swift}-BAT 100-month catalog \citep{Segreto:2015pk} contains 1710 hard X-ray sources and has been used extensively for extended campaigns of follow-up observations with X-ray facilities aimed at obtaining a complete characterization of the obscuring medium surrounding accreting SMBHs at $z<0.05$ \citep[e.g.,][]{2019ApJ...872....8M,2019ApJ...871..182Z,2019ApJ...870...60Z,2021ApJ...922..252T,2021A&A...650A..57Z,2022ApJ...940..148S,2023A&A...676A.103S}. Many studies at different wavelengths \citep[e.g.,][]{1985ApJ...297..621A, 2019A&A...623A..79C, 2019ApJ...884..171H} across the years have allowed us to build a scenario where most of this obscuration is due to a distribution of material located in close proximity (1 - 10 pc) of the accreting supermassive black hole, historically called the torus. The atomic and molecular gas density in the torus
is typically parameterized in X-rays as the line-of-sight neutral hydrogen column density $N_{H,los}$. When the obscuring matter has a column density equal to or greater than the inverse of the Thomson cross section $\sigma_{T}$, $N_{H,los}\ge \sigma_{T}\simeq 1.5\times10^{-24}$ $cm^{-2}$, the source is defined as Compton thick (CT--) AGN. The presence of CT-AGN can explain the excess of emission around the Compton hump ($\sim$ 30 keV) in the CXB. Models of AGN population indicate that the Compton-thick fraction in the local universe could be up to 30--50\% \citep[e.g.,][]{2007A&A...463...79G,2019ApJ...871..240A}. However, X-ray surveys so far have shown that this fraction is only $<$ 10\% \citep{ricci2015compton,2021ApJ...922..252T}, indicating a population of CT-AGN still undetected. A hard X-ray survey is thus essential in finding CT-AGN candidates in this scenario, considering that the highly energetic hard X-ray photons can pass through  large column densities of dust and gas, revealing CT-AGN \citep[e.g.,][]{2008ApJ...685L..19B,ricci2015compton,koss2016new}.

The \textit{Swift}-BAT 150-month catalog (Segreto et al. in prep. available online at \href{https://science.clemson.edu/ctagn/bat-150-month-catalog/}{Swift-BAT 150-month catalog}) comprises 2339 sources, of which 344 lack a low-energy counterpart. Among these 344 sources, 
 
250 have available soft (i.e., at energies below 10 keV) X-ray data. For those 250 sources, there are 251 possible counterparts, among which 179 counterparts are above the galactic plane ($|b| > 10^\circ$) and 72 counterparts are within the galactic plane ($|b| < 10^\circ$). In this work, we use these data to determine counterparts and present detailed source classifications for 250 sources.

Our work is organized as follows: Section \ref{sec:data analysis} discusses the data analysis of soft X-ray observations. Section \ref{sec:procedure} is devoted to the procedure of counterpart association for BAT sources. Section \ref{sec:Properties of the BAT 150-month catalog formerly unassociated sources} presents the properties of the 250 BAT sources. Finally, Section \ref{sec:Result} is dedicated to the results of our work and further discussion.

We use \cite{1989GeCoA..53..197A} cosmic abundances, fixed to the solar value, and the \cite{1996ApJ...465..487V} photoelectric absorption cross-section. The luminosity distances in this work are computed assuming a cosmology with $H_{0}$ = 70 km $s^{-1}$ $Mpc^{-1}$, ${\Omega}_{M}$= 0.3 and ${\Omega}_{\Lambda}$= 0.7.

\section{Data analysis of soft X-ray observations} \label{sec:data analysis}
 We submitted\footnote{PI: Torres-Alba; proposal ID: 1821052} (and were granted) a \textit{Swift} proposal to observe for 2\,ks, with \textit{Swift}-XRT, 111 (out of 344) 150-month \textit{Swift}-BAT sources  lacking soft X-ray coverage. For the remaining 233 unassociated BAT sources, we search for soft counterparts by employing data from the X-ray Multi-Mirror Mission \citep[\textit{XMM-Newton}][]{2001A&A...365L...1J}, \textit{Chandra} X-ray telescope \citep{2002PASP..114....1W}, \textit{Swift}-XRT \citep{2005SSRv..120..165B} and extended ROentgen Survey with an Imaging Telescope Array \citep[{\textit{eROSITA}}][]{2007SPIE.6686E..17P,2021A&A...647A...1P} archives, and find observations for 139 out of 233 sources. For the remaining 94 sources, either there is no observation available, or there is 
some X-ray observation covering the BAT position (by \textit{Swift}-XRT), but no sources are detected over the background.

 For each source, we download the \textit{XMM-Newton}/ \textit{Chandra}/ \textit{Swift}-XRT data from NASA HEASARC (High Energy Astrophysics Science Archive Research Center\footnote{\url{https://heasarc.gsfc.nasa.gov/cgi-bin/W3Browse/w3browse.pl}}) 
 querying around the BAT source coordinates with the maximum allowed separation radius of 6$\arcmin$ between the BAT coordinates and the center of the X-ray observation. We find \textit{XMM-Newton} observations for 10 sources. We reduce the \textit{XMM-Newton} data using the SAS v18.0.0 \citep{2017xru..conf...84G} after cleaning for flaring periods, adopting standard procedures. We find \textit{Chandra} data for 13 sources. We reduce the \textit{Chandra} data using CIAO v4.12 \citep{2006SPIE.6270E..1VF}. We then find 201 sources which have been observed with \textit{Swift}-XRT. We reduce the \textit{Swift}-XRT data using HEASoft version 6.31 \citep{2014ascl.soft08004N}. In cases where multiple XRT observations are available,
 we stack the event files into one summed event file following the XSELECT user guide. Similarly, we stack images using the XIMAGE instruction manual for \textit{Swift}-XRT. We use the XRT instrument software version 3.7.0 for analysis with the most recent CALDB version at the time of analysis (1.0.2). Finally, we search for counterparts in the \textit{eROSITA} \citep{2024A&A...682A..34M} catalog, and find a counterpart for 27 BAT sources lacking a soft X-ray counterpart from any of the other instruments mentioned before. These sources are within the region of the western Galactic hemisphere, covered by the public \textit{eROSITA} data release. For each source, we use the \textit{eROSITA} sky view tool\footnote{\href{https://erosita.mpe.mpg.de/dr1/erodat/skyview/sky/}{https://erosita.mpe.mpg.de/dr1/erodat/skyview/sky/}} \citep{2024A&A...682A..35T} to identify counterparts within the BAT $R_{95}$ region.

\subsection{Association of \textit{Swift}--BAT sources with multiple soft X-ray counterparts}
There are 30 cases in \textit{Chandra} and \textit{Swift}-XRT observations where multiple soft X-ray sources (0.3--10 keV) exist within the BAT 95\% confidence positional uncertainty region (BAT $R_{95}$). In such cases, a spectrum is extracted to compare the two possible counterparts. For \textit{Chandra}, we extract the source spectra from a {5\arcsec} circular region centered around the source, and the background spectra are obtained using an annulus (inner radius 6\arcsec, outer radius 15\arcsec) surrounding the source, excluding any resolved sources. For \textit{Swift}-XRT observations, the source spectra are extracted from a {20\arcsec} circular region, while the background spectra are obtained using an annulus (inner radius 30\arcsec, outer radius 60\arcsec) surrounding the source, excluding any resolved sources. We then fit the data using XSPEC \citep{1996ASPC..101...17A} version 12.13.0 with a power-law model and employing c-statistics \citep{1979ApJ...228..939C}. We further use a power-law model to extrapolate the observation data to the 15--150 keV range to estimate the flux for all sources within the $R_{95}$. We consider only the brightest source (i.e., the source with the highest 15--150 keV flux) as the possible counterpart if the difference between extrapolated fluxes is larger by a factor of 10. We further discuss the counterpart identification process for these BAT sources in Section~\ref{sec: flag s and flag m}.

\section{Procedure for source association} \label{sec:procedure}

We use soft X-ray observations to identify counterparts for BAT-detected sources because telescopes covering the 0.5--10 keV energy range typically offer much better positional accuracy than hard X-ray missions. The on-axis point spread function (PSF) is usually described by the half energy width (HEW), defined as the diameter containing 50\% of the flux. For XMM-Newton, at 1.5 keV, the HEW values are {16.6\arcsec} for pn, {16.8\arcsec} for MOS1, and {17.0\arcsec} for MOS2\footnote{\href{https://xmm-tools.cosmos.esa.int/external/xmm_user_support/documentation/uhb/onaxisxraypsf.html}{https://xmm-tools.cosmos.esa.int/external/xmm\_user\_support
/documentation/uhb/onaxisxraypsf.html}}. For Chandra, the on-axis HEW  
is less than {0.5\arcsec} (Chandra Proposal Guide v.15\footnote{\href{https://cxc.harvard.edu/cdo/about\_chandra/}{https://cxc.harvard.edu/cdo/about\_chandra/}}). For \textit{Swift}-XRT, the HEW values measured on-axis range from {16\arcsec} (at 0.28 keV) to {22\arcsec} (at 8.05 keV) \citep{2004SPIE.5165..232M}. The on-axis HEW for \textit{eROSITA} is {15\arcsec}  \citep{2010AIPC.1248..543P}. On the other hand, the median BAT 95\% confidence positional uncertainty region ($R_{95}$) is {4.3\arcmin} (see Section \ref{sec:Result}).

In most soft X-ray observations, the field of view (FoV) typically includes a single soft X-ray source within the BAT $R_{95}$ region, which we identify as the likely counterpart. Previous instances of the 54--month Palermo BAT catalog \citep{2010A&A...524A..64C} use a {6\arcmin} radius\textbf{\footnote{However, the \textit{Swift}-BAT 157-month catalog adopts a {12\arcmin} radius to identify counterparts.}} \citep[99.7\% confidence level for a source detection at 4.8$\sigma$][]{2010A&A...510A..47S}. Therefore, we consider (as the counterpart of the BAT source) those soft X-ray sources located at a distance between $R_{95}$ and 6\arcmin, if no soft X-ray source is found within $R_{95}$.

Using the coordinates of the soft X-ray counterpart, we attempt to associate the source with a lower frequency counterpart whenever possible, in doing so providing the object type and redshift. Whenever feasible, we report the source based on its type (e.g., galaxy, AGN and blazar) from the SIMBAD\footnote{\href{https://simbad.cds.unistra.fr/simbad/sim-fcoo}{https://simbad.cds.unistra.fr/simbad/sim-fcoo}}/NED\footnote{\href{https://ned.ipac.caltech.edu/}{https://ned.ipac.caltech.edu/}} databases. If the source type classification is not available on the SIMBAD/NED databases, we report the source as infrared (IrS), radio (RadioS), or soft X-ray (X, energy band 0.3--10 keV).

In the following subsections, we present a set of flags to allow the reader to understand how the counterpart is chosen. The different flags are summarized in table \ref{tab:flags}.

\subsection{Flag 1 and Flag 2}
\label{sec: flag 1 and flag 2}
\textit{Flag 1}: it indicates cases where the candidate counterparts (single or multiple sources) are within the respective $R_{95}$. Out of 251 associated BAT sources, we have 197 (78\%) sources with Flag 1.

\textit{Flag 2}: it identifies sources that do not have a counterpart within the $R_{95}$ circle but have one within a {6\arcmin} radius. There are 48 sources located in the region between the $R_{95}$ and a {6\arcmin} radius. Additionally, 6 sources are detected with their emission centroid outside the {6\arcmin} circle, but their localization uncertainty region overlaps with the {6\arcmin} circle. These sources are also classified under Flag 2. In total, out of 251 associated BAT sources, we have 54 (22\%) sources categorized under Flag 2.
\subsubsection{Flag s and Flag m}
\label{sec: flag s and flag m}
\begin{figure}[ht]
    \includegraphics[width=0.46\textwidth]{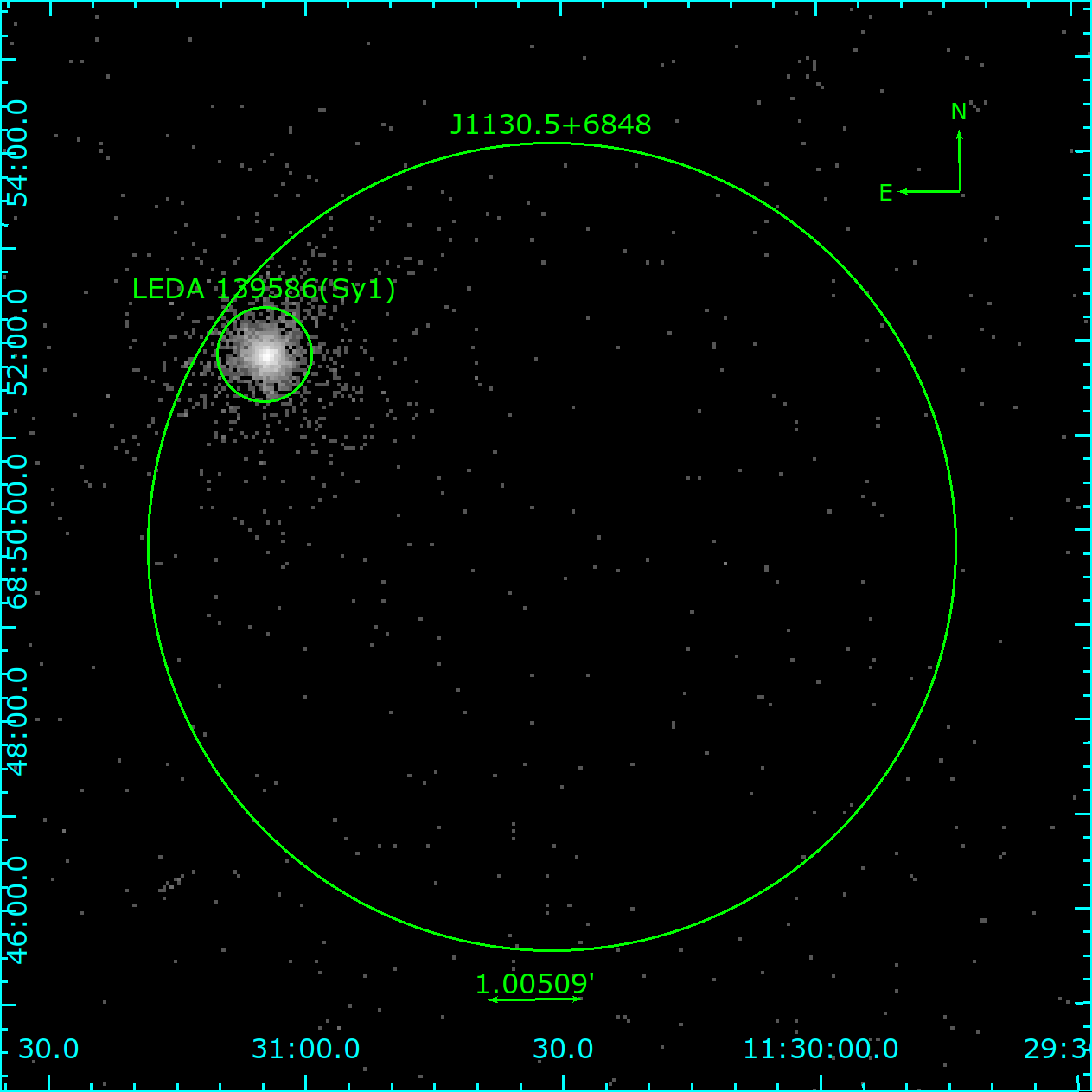}
    \caption{\textit{Swift}-XRT observation of a single source (smaller circle) in the BAT $R_{95}$ (bigger circle) for the BAT source J1130.5+6848 in the energy band 0.3--10 keV. Only one bright soft X-ray source (whose optical counterpart is the galaxy LEDA 139586) is present within the BAT $R_{95}$, which we associate as the BAT counterpart.}
    \label{fig:single source}
\end{figure}

Source with single soft X-ray counterpart candidate (s): when only one soft X-ray source is present within the BAT $R_{95}$, we assume that source as the counterpart to the BAT detection. Figure \ref{fig:single source} illustrates such a case, where only one soft X-ray source is clearly found within the BAT $R_{95}$.

Source with multiple soft X-ray counterpart candidates (m): in this scenario, multiple sources are present within the BAT $R_{95}$. We extract a spectrum for each source individually and then analyze that using XSPEC as detailed in Section \ref{sec:data analysis}, and further extrapolate the data using a power-law model to predict flux in the 15--150 keV band; we assume the source with higher flux as the BAT counterpart. 

In the case of flag `m', we mostly fit unobscured power-laws, but we encountered sources for which we had to fit an obscured power-law. We did so, when an unobscured power-law resulted in a poor fit (i.e. cstat/d.o.f $>$ 1.5), or the photon index of a simple power-law was $<$1.4. Depending on data quality, we fit both the photon index and the absorbing column density; for those sources without enough counts ($<$ 55 counts) in the 0.3--10 keV band, we instead chose to fix the photon index to the typical value, for Seyfert galaxies, of $\Gamma$=1.8 \citep[e.g.,][]{Dadina2008,Ricci2011}, and measured the line-of-sight column density by leaving it as a free parameter.

\begin{figure*}[ht]
    \centering
    \includegraphics[width=1\textwidth]{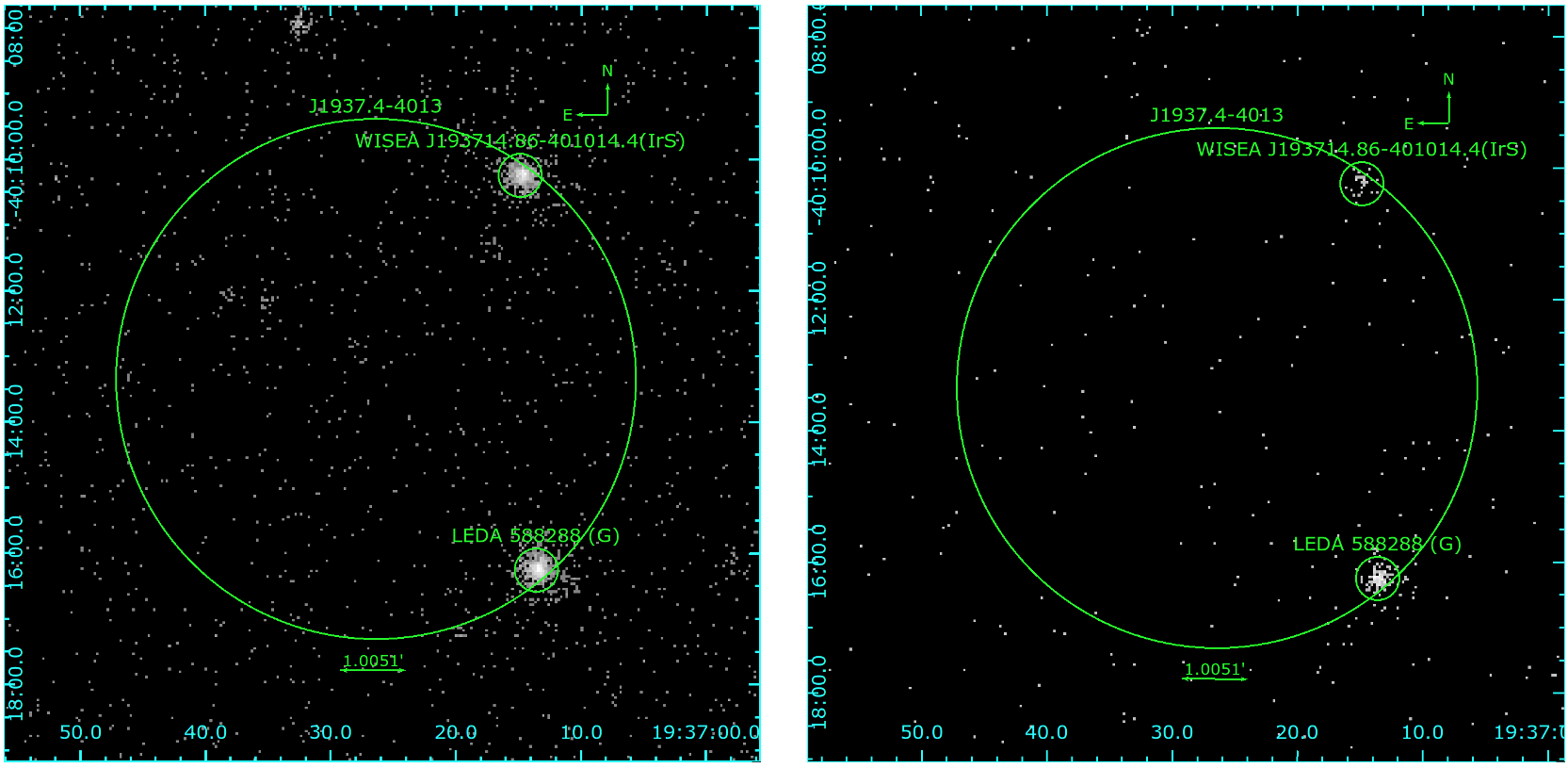}
    \caption{\textit{Swift}-XRT observation for the BAT source J1937.4-4013 in the energy band 0.3--5 keV (left panel) and 5.1--10 keV (right panel). There are two soft X-ray sources within the BAT $R_{95}$, WISEA J193714.86--401014.4 and LEDA 588288 with flux 1.2 $\times$ 10$^{-12}$ erg s$^{-1}$ cm$^{-2}$ and 6.5 $\times$ 10$^{-12}$ erg s$^{-1}$ cm$^{-2}$ respectively in the energy band 0.3--10 keV. The spectral fitting and power-law extrapolation predicts a flux of 1.1 $\times$ 10$^{-12}$ erg s$^{-1}$ cm$^{-2}$ and 6.1 $\times$ 10$^{-10}$ erg s$^{-1}$ cm$^{-2}$ for WISEA J193714.86--401014.4 and LEDA 588288 respectively in the energy band 15--150 keV, so we associated LEDA 588288 to the BAT source.}
    \label{fig:multiple}
\end{figure*}

Figure \ref{fig:multiple} illustrates a case where two sources are present in the BAT $R_{95}$ in the 0.3--10 keV band,  WISEA J193714.86--401014.4 and LEDA 588288. The spectral fitting and power-law extrapolation predict a flux in the range 7.3$\times$10$^{-13}$ erg s$^{-1}$ cm$^{-2}$ to 1.8$\times$10$^{-12}$ erg s$^{-1}$ cm$^{-2}$ and 3.8$\times$10$^{-10}$ erg s$^{-1}$ cm$^{-2}$ to 8.5$\times$10$^{-10}$ erg s$^{-1}$ cm$^{-2}$ for WISEA J193714.86--401014.4 and LEDA 588288 respectively in the energy band 15--150 keV, we thus assumed LEDA 588288 as the BAT counterpart.

For just one source in our sample, J1159.5+2913, we end up having a double association. Two soft X-ray sources are present in the $R_{95}$; Ton 599 (Seyfert 1; Sy1 hereafter) at redshift 0.72  with flux 2.4 $\times$ 10$^{-12}$ erg s$^{-1}$ cm$^{-2}$ in the 0.3--10 keV band and projected flux of 5.1 $\times $10$^{-12}$ erg s$^{-1}$ cm$^{-2}$ in 15--150 keV band) and LEDA 1857480 (Sy1 at redshift 0.083 with flux 1.1 $\times$ 10$^{-12}$ erg s$^{-1}$ cm$^{-2}$ in the 0.3--10 keV band and projected flux of 8.0 $\times$ 10$^{-12}$ erg s$^{-1}$ cm$^{-2}$ in 15--150 keV band). As they have a similar extrapolated flux in the 15--150 keV band, we do not pick one over the other as the most likely counterpart, and in fact we suspect they both are likely contributing to the BAT source.

\subsubsection{Flag eR}
\textit{eROSITA source (eR)}: these sources are detected using \textit{eROSITA} Data Release 1 \citep{2024A&A...682A..34M,2024A&A...682A..35T}. We employ the \textit{eROSITA} Sky View Tool to identify soft X-ray sources within the BAT $R_{95}$. If a single source is found within the $R_{95}$, we classify it as the potential counterpart. In cases where multiple \textit{eROSITA} sources are detected, we designate the source with higher flux in the 0.2--2.3 keV energy range as the primary counterpart. If multiple sources exhibit similar fluxes (flux in the order of $10^{-13}$ erg s$^{-1}$ cm$^{-2}$ in the energy band 0.2--2.3 keV), we refrain from designating any of them as a possible counterpart in this catalog due to the lack of an effective criterion for selection. We assign a separate flag to \textit{eROSITA} sources as further scrutiny is needed to confirm those,  due to the soft nature of the \textit{eROSITA} catalog. We recommend to exercise caution when taking any of these sources as the true BAT counterpart, particularly in cases where multiple candidates are present. 

\subsection{Unassociated and Unobserved}

\textit{Unassociated sources (ua)}: these are the sources for which soft X-ray observation data are available, but no soft X-ray source is found within the BAT $R_{95}$ or 6$\arcmin$ in the 0.3--10 keV band. There are 57 such sources; for each, we provide the BAT-detected coordinates in Table \ref{tab:all_observ}.

\textit{Unobserved sources (uo)}: there are 37 sources for which either no soft X-ray observation is available or the BAT-detected region has been only partially observed. Therefore, we report them as unobserved and provide their BAT-detected coordinates in the Table \ref{tab:all_observ}.

\begin{table}[ht]
\caption{Summary of symbols used to flag different source associations.\label{tab:flags}}
\centering
\begin{tabular}{cc}
\hline\hline
Flag & Meaning \\
\hline\hline
1 & Counterpart within $R_{95}$ \\
2 & Counterpart in the range $R_{95}-6'$ \\
\hline
s & Only one potential counterpart \\
m & Multiple potential counterparts \\
\hline
eR & Counterpart associated using eROSITA \\
\hline\hline
ua & \begin{tabular}{@{}c@{}}Unassociated \\ (i.e. data available, but no counterpart detected)\end{tabular} \\
\hline\hline
uo & Unobserved \\
\hline\hline
\end{tabular}
\tablefoot{ Flags within the same horizontal box are complementary (i.e. a source cannot be flagged 1 and 2 at the same time, but can be flagged, e.g., 1s). However, sources flagged as `ua' or `uo' cannot have any other associated flag. A more in-depth explanation can be found in Sect.~\ref{sec:procedure}.}
\end{table}

\section{Description of the newly associated BAT sources} \label{sec:Properties of the BAT 150-month catalog formerly unassociated sources}

We present a catalog of 250 originally unassociated sources from the \textit{Swift}-BAT 150-month catalog (Segreto et al., in preparation). We list the first few sources from the catalog in Table \ref{tab:all_observ} (the full catalog is attached as a machine readable file), which includes the following information:

\begin{itemize}
    \item Col. 1 reports the BAT catalog name of the source.
    \item Col. 2 and Col. 3 represent right ascension (RA BAT) and declination (Dec BAT) of the BAT source in the FK5 coordinate system.
    \item Col. 4 reports the BAT 95\%  confidence positional uncertainty region ($R_{95}$) in arcmin.
    \item Col. 5 reports the flags, as presented in Section \ref{sec:procedure}.
    \item Col. 6 and Col. 7 report the RA and Dec of the counterparts, determined from the SIMBAD/NED databases.
    \item Col. 8 and Col. 9 reports the galactic latitude (l) and galactic longitude (b) of the counterparts.
    \item Col. 10 reports the name of the counterpart, as taken from the SIMBAD/NED databases.
    \item Col. 11 reports the source type of the counterpart, as reported in the SIMBAD/NED databases. This includes galaxies with reported AGN activity like Seyfert galaxies of unknown types (SyG), Seyfert 1 (Sy1), Seyfert 2 (Sy2), Quasars (QSO), Blazar (Bz) and galaxies without any reported AGN activity (G). This also includes sources not firmly associated to an object type, and instead classified as infrared infrared (IrS), radio (RadioS) and soft X-ray sources (X).
    \item Col. 12 reports the redshift taken from the SIMBAD/NED databases. We report sources with no available redshift as N/A.
\end{itemize}

\begin{center}
\begin{table*}[ht]
\centering
 \caption{Observational details and properties of a fraction of the sources analyzed in this work (refer to Table \ref{Counterpart types} for a list of objects types). This table is available in machine readable format. }
 \label{tab:all_observ}
\vspace{.1cm}
\renewcommand*{\arraystretch}{1.5}
\resizebox{1\textwidth}{!}{%
  \begin{tabular}{cccccccccccc}
    {BAT Name} & {RA BAT} & {Dec BAT}& {$R_{95}$}  & {Flags} & {RA OBJ} & {Dec OBJ} & {Galactic l} & {Galactic b} &{Counterpart} & {Type} & {Redshift} \\
  & (J2000) & (J2000) & (arcmin) &  & (J2000) & (J2000) &  & \\ 
 \hline\hline
 J0001.5+1113 & 0.395 & 11.217 & 4.5 & 2s & 0.461 & 11.280 & 103.911 & -49.8 & 2MASS J00015055+1116471 & Sy1 & 0.158 \\ 
 \hline
 J0007.3-4121 & 1.863 & -41.367 & 4.5 & 1s & 1.778 & -41.356 & 332.708 & -73.1 & ESO 293-37 & Sy2 & 0.047 \\
 \hline
 J0022.2+8042 & 5.527 & 80.730 & 4.2 & 1s & 5.682 & 80.729 & 121.719 & 17.9 & 2MASX J00224371+8043462 ID & Sy1 & 0.074 \\
 \hline
 J0027.6+2842 & 6.893 & 28.711 & 3.8 & 1s & 6.897 & 28.708 & 116.632 & -33.9 & RX J0027.5+2842 & AGN & 0.063 \\
 \hline
 J0030.4-5303 & 7.613 & -53.063 & 4.4 & 1s & 7.682 & -53.105 & 309.965 & -63.7 & WISEA J003043.69-530617.4 & IrS & N/A \\
 \hline
 J0039.9-1553 & 9.975 & -15.884 & 4.2 & ua & 9.975 & -15.884 & 108.95 & –78.44 & J0039.9-1553 & X & N/A \\
 \hline
 J0047.4+5449 & 11.845 & 54.831 & 4.9 & uo & 11.845 & 54.831 & 122.34 & –8.04 & J0047.4+5449 & X & N/A \\
 \hline
 J0051.8+4418 & 12.975 & 44.295 & 3.9 & 2m & 13.057 & 44.331 & 123.081 & -18.5 & UGC 530 & SyG & 0.017 \\
 \hline
 J0058.3-6652 & 14.575 & -66.875 & 4.9 & 2m & 14.683 & -66.883 & 301.813 & -50.2 & LEDA 301110 & G & 0.085 \\
 \hline
 J0059.8+3911 & 14.964 & 39.195 & 3.9 & 1m & 14.934 & 39.171 & 124.688 & -23.7 & 2MASX J00594418+3910139 & G & 0.053 \\
 \hline
 J0107.7+5745 & 16.912 & 57.782 & 4.8 & uo & 16.912 & 57.782 & 125.10 & –5.02 & J0107.7+5745 & X & N/A \\
 \hline
 J0113.8+2519 & 18.425 & 25.328 & 3.9 & 2s & 18.345 & 25.315 & 129.167 & -37.3 & SDSS J011322.69+251853.2 & QSO & 1.589 \\
 \hline
  ... & ... & ... & ... & ... & ... & ... & ... & ... & ... \\  
 \hline    
   \end{tabular}
}
\vspace{.2cm}
\end{table*}
\end{center}

\section{Results and discussion} \label{sec:Result}

As discussed in the previous sections, for the complete sample of 344 initially unassociated sources, soft X-ray observations are available for 250 sources. Using the association methods described in Section \ref{sec:procedure}, we identify possible counterparts for those sources (see Table \ref{tab:all_observ}). For the remaining sources, data are either absent or insufficient, which means the BAT $R_{95}$ is covered partially, or the exposure time is too short (typically $<$ 0.7 ks) to confirm the counterpart. We note that among the 251 identified sources, 179 sources lie outside the Galactic plane ($|b| > 10\degree{}$) and 72 sources lie within it ($|b| < 10\degree{}$), as shown in Table \ref{Counterpart types}. 

\begin{table*}[ht]
\centering
\caption{Counterpart types of BAT sources.\label{Counterpart types}}
\begin{tabular}{cccc}
\hline\hline
Source type & N.src. $|b| > 10^\circ$ & N.src. $|b| < 10^\circ$ & Total \\
\hline\hline
Seyfert I (Sy1) & 34 & 2 & 36 \\
Seyfert II (Sy2) & 32 & 3 & 35 \\
Seyfert galaxy of unknown type (SyG) & 2 & 0 & 2 \\
Active Galactic Nuclei (AGN) & 7 & 0 & 7 \\
Quasar (QSO) & 9 & 0 & 9 \\
Blazar (Bz) & 1 & 0 & 1 \\
Blazar candidate (Bz?) & 1 & 0 & 1 \\
BL Lac & 2 & 1 & 3 \\
Galaxies (G) & 55 & 3 & 58 \\
Soft X-ray sources (X) & 8 & 36 & 44 \\
Infrared source (IrS) & 20 & 18 & 38 \\
Pulsars (PSR) & 0 & 2 & 2 \\
Globular Cluster (*Cl) & 0 & 1 & 1 \\
Cataclysmic Variable (CV) & 1 & 2 & 3 \\
Cataclysmic Variable candidate (CV?) & 0 & 1 & 1 \\
Young Stellar Object (Y*O) & 0 & 1 & 1 \\
Young Stellar Object Candidate (Y*?) & 1 & 0 & 1 \\
Radio source (RadioS) & 5 & 1 & 6 \\
Gamma-ray source (GammaS) & 1 & 1 & 2 \\
\hline
Total & 179 & 72 & 251 \\
\hline
\end{tabular}
\tablefoot{ N.src. = Number of sources. \newline We note that GammaS is used to stay true to the SIMBAD classification, but actually refers to INTEGRAL-detected sources.}
\end{table*}

\begin{figure}[ht]
    \centering
    \includegraphics[width=0.45\textwidth]{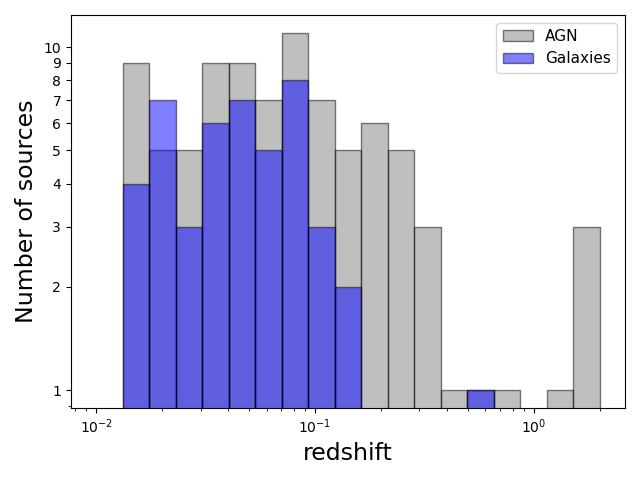}
    \caption{Redshift distribution of 135 BAT detected sources for which a redshift information was obtained thanks to the associations provided by our analysis. The AGN sources includes Sy1/Sy2/SyG, AGN and QSO, whereas the Galaxies are galaxies in which AGN activity has not been reported until now.}
    \label{fig:Redshift distribution}
\end{figure}

Out of 251 sources, redshift information is available for 139 sources on the SIMBAD/NED databases or the Milliquas catalog \citep{2023OJAp....6E..49F}. Among these 139 sources, 98 are located at $z<0.1$, and 37 additional sources\footnote{We exclude the two BL Lacs, one known blazar and one blazar candidate in our sample from the redshift range computation.} are in the range $0.1<z<1.9$. Of the 135 sources that are not blazars, 89 are known AGN (including Seyferts, AGN, and QSOs), and 46 are classified as galaxies (galaxies with no prior knowledge of the presence of an AGN). Figure \ref{fig:Redshift distribution} shows the redshift distribution of these 89 AGN and 46 galaxies.

\begin{figure*}[ht]
    \centering
    \includegraphics[width=18cm]{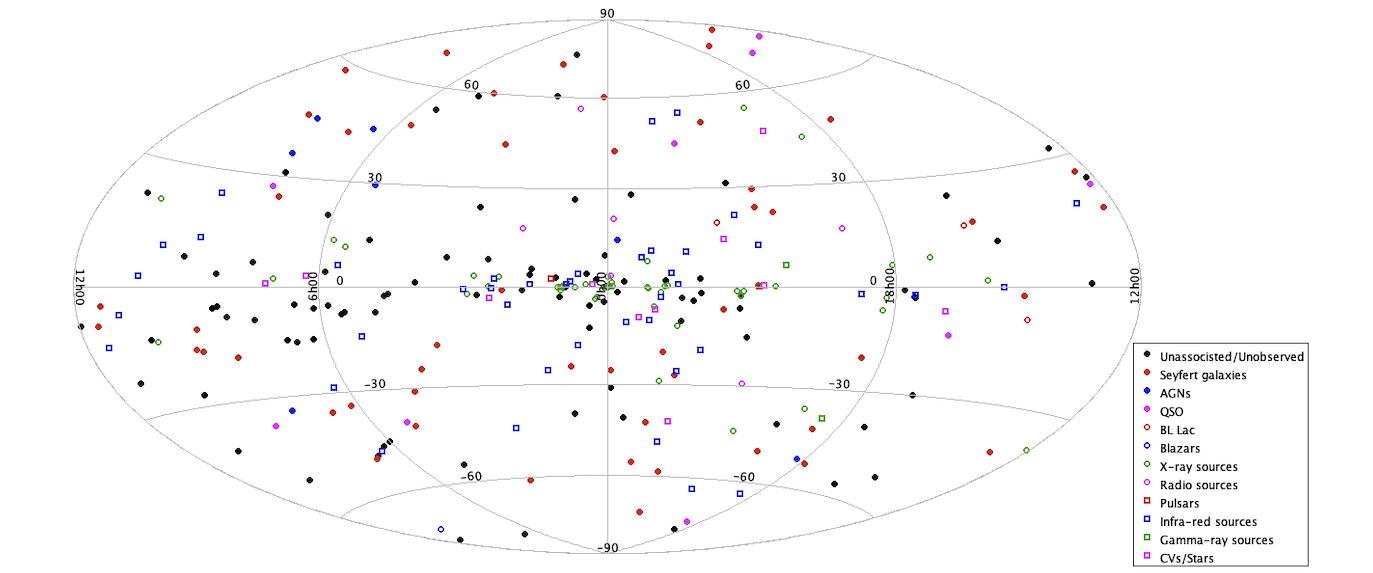}
    \caption{All-sky map showing the classification of the 344 sources from BAT 150-month survey sources. The figure uses a Hammer-Aitoff projection in Galactic coordinates.}
    \label{fig:skymap}
\end{figure*}

Figure \ref{fig:skymap} presents the distribution of sources in Galactic coordinates using a Hammer-Aitoff projection. It is observed that many sources categorized as X-ray sources (X), infrared sources (IrS), or those that remain unassociated/unobserved and lack definitive identification, are located above the Galactic plane. This suggests that a non-negligible fraction of the unassociated sources are in fact AGN. Further investigation of these sources may confirm, or rule out, this hypothesis.

\subsection{Comparison to previous catalogs}

\begin{figure}[ht]
    \centering
    \includegraphics[width=0.47\textwidth]{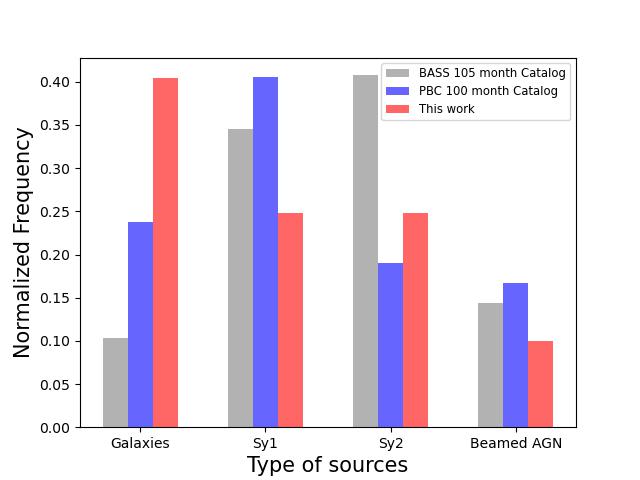}
    \caption{A normalized source type comparison with PBC 100 month catalog and BASS 105 month catalog. The beamed AGN includes blazars, FSRQ and QSO. The galaxies are classified as unknown AGN in the BASS 105 month catalog. Based on their X-ray luminosity, we can safely claim that all the sources optically classified as galaxies actually host an AGN.}
    \label{fig:Redshift}
\end{figure}

Figure \ref{fig:Redshift} compares the normalized source type distribution of counterparts identified in our analysis, with that of the PBC 100-month catalog and the BASS 105-month catalog. 

As can be seen, this work finds a much higher incidence of ``normal'' galaxies ($>40$\%, compared to $\sim10-20$\%) than previous catalogs. It is worth mentioning, however, that all 58 sources (see Table \ref{Counterpart types}) that are detected by BAT and are classified as normal galaxies on the SIMBAD/NED databases are actually likely to be AGN. This is because BAT detects sources with high X-ray luminosities ($>10^{42}$ erg s$^{-1}$) in a very hard energy range (15--150 keV). Normal galaxies generally do not have emission mechanisms capable of producing detectable X-ray emission at energies above 15\,keV (see e.g.  \citealt{2021ApJS..257...61Y} and \citealt{2021MNRAS.506.5935R}), where even \textit{NuSTAR} struggles to detect extreme star-forming galaxies at $<$ 20 keV, in the absence of an AGN). This is further supported by the fact that we detect a much lower incidence of Sy1 ($\sim25$\% of sources, compared to $\sim35-40$\%) and Sy2 galaxies ($\sim25$\% of sources, compared to $\sim40$\% in the BASS 105 month catalog). Interestingly, the PBC 100 month catalog suffers a similar issue between ``normal'' galaxies and Sy2s, which may be due to the BASS collaboration's efforts of follow-up observations, which are likely responsible for the optical classification of a significant fraction of those sources \citep[e.g.][]{Lamperti2017,Koss2022,Oh2022}.

Previous studies \citep{Segreto:2015pk,2018ApJS..235....4O} indicate that such BAT-detected ``normal'' galaxies, when indeed found to be AGN, tend to be more obscured than average galaxies. This could also be the reason some of them are missed or misclassified in surveys, which happens less frequently in Sy1 galaxies. As such, all sources listed as galaxies in this catalog are valuable targets for follow-up observations and, in particular, searches for highly obscured (and even Compton-thick) AGN \citep{2016A&A...594A..73A,2019ApJ...872....8M,2022ApJ...940..148S,2023A&A...676A.103S,2023RAA....23e5002M}.

We note that we only compare galaxies, Seyferts, and Blazars to previous catalogs, but this work contains a significant number of counterparts classified as Infrared ($\sim17$\%) and soft X-ray ($\sim$18\%) sources. Previous catalogs do not have significant populations of counterparts without an associated type, and thus we have not added these to the comparison. As can be seen in Table \ref{Counterpart types}, the majority of these sources fall within the Galactic plane, making it impossible to determine their source type without a spectroscopic follow-up campaign. The 28 sources that fall outside of the plane, however, are likely to be AGN. The 5 radio sources outside the Galactic plane are most likely to be blazars.

\begin{figure}[ht]
    \centering
    \includegraphics[width=0.47\textwidth]{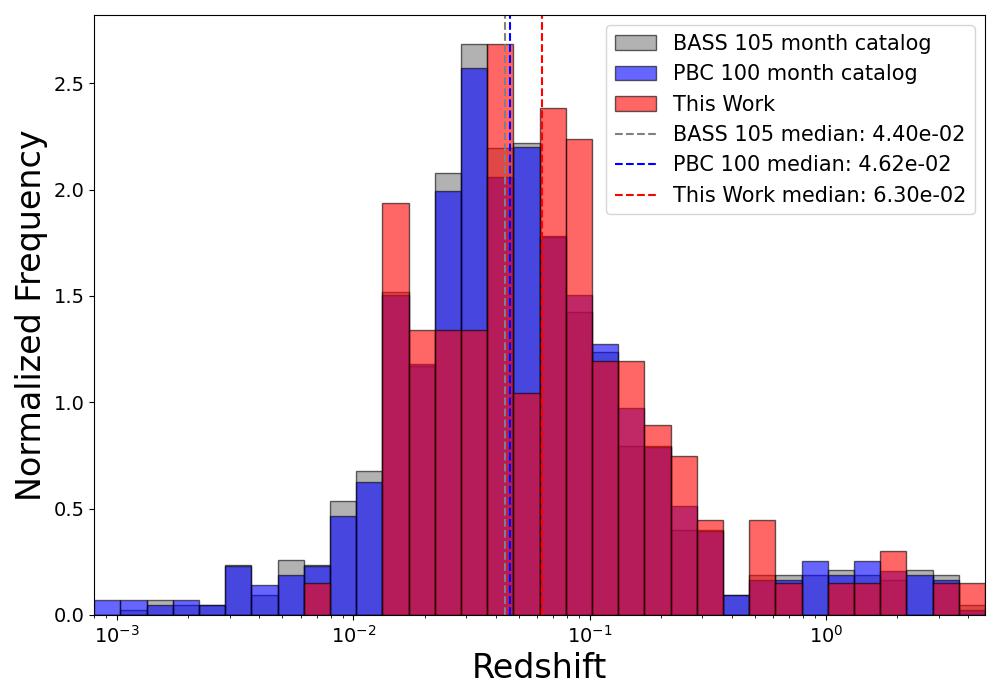}
    \caption{A normalized comparison of redshift distribution for 135 out of 251 BAT detected sources from this work with 986 sources from previous PBC 100 month catalog and with 966 sources from previous BASS 105 month catalog. The redshift distribution is similar to PBC 100 month catalog and BASS 105 month catalog showing the accuracy of this work. }
    \label{fig:Redshift100mcomparison}
\end{figure}

Figure \ref{fig:Redshift100mcomparison} presents a comparison between the normalized redshift distribution of the  BAT AGN Spectroscopic Survey 105-month catalog \citep[BASS 105-month catalog;][]{2018ApJS..235....4O}, the Palermo BAT 100-month catalog \citep[PBC 100-month catalog;][]{Segreto:2015pk} and our study. The median redshift of the BASS 105 month catalog distribution is 0.044, and the lower (2.5th) and upper (97.5th) percentiles are 0.005 and 1.83 respectively. The median redshift of the PBC 100-month catalog distribution is 0.046, and the lower and upper percentiles are 0.003 and 1.64, respectively. In this work, the median redshift of the distribution is 0.063, and the lower and upper percentile values are 0.0154 and 1.75, respectively. 

This comparison shows that the median redshift of our sample is slightly higher than that of previous works, which is reasonable as the sources in our sample are fainter (since they were undetected within the first 100 months of BAT observations, but detected with an additional 50 months of observation) than the ones in the BASS 105-month catalog and PBC 100-month catalog. Since in the 15--150\,keV energy band the effect of $N_{H}$ on the observed flux is marginal, we can assume that a fainter flux is related to a (slightly) higher redshift, or to an intrinsically lower luminosity. Table \ref{Meanmedian} also shows the mean and median redshift of the catalogs. 
\begin{table}
\caption{ Mean and Median of redshift distribution. \label{Meanmedian}}
\begin{tabular}{ccc}
Catalog & Mean redshift & Median redshift \\  
 \hline\hline
 BASS 105 month  & 0.173$\pm$ 0.015 & 0.044$\pm$ 0.002 \\
 PBC 100 month & 0.170$\pm$ 0.014 & 0.046$\pm$ 0.002  \\
 This work & 0.190$\pm$ 0.043 & 0.063$\pm$ 0.007 \\
 \hline
\end{tabular}
\end{table}

 The most recent \textit{Swift}-BAT 157-month \citep{2025arXiv250604109L} catalog contains 1888 sources, including 256 new detections above the $4.8\sigma$ significance threshold. In that catalog, the BAT $R_{95}$ positional uncertainty extends up to 12\arcmin, whereas in this work, we adopt an uncertainty radius of up to $6\arcmin$. A cross-match between the selected counterparts
(with $20\arcsec$ max error) in this work and the counterparts to BAT sources in the \textit{Swift}-BAT 157-month catalog identifies 48 counterparts in common. If, instead, we cross-match against the BAT source centroid and use the maximum error of 12 \arcmin (as considered by the 157-month catalog), we get 114 matches. That is, 114 of our counterparts could be counterparts to sources in the \textit{Swift}-BAT 157-month catalog.

Out of these 114 sources:
\begin{itemize}
    \item 39 sources do not have a counterpart in the 157-month catalog, but have a counterpart in this work.
    \item 4 sources are unknown (no counterpart) in both the 157-month catalog and this work. In this case, the coordinates we are using to cross-match are of our BAT source detection, and not those of a selected counterpart.
    \item  9 sources have counterparts in the 157-month catalog but do not have one in our work (among those, 8 sources are flagged as unassociated (ua) and 1 source is flagged as unobserved (uo) in this work). We attribute this to two main reasons: 1) Some of the counterparts proposed in the 157-month catalog are at a distance  $>$ $7 \arcmin$ from the BAT source centroid, while we consider a maximum $R_{95}$ up to $6 \arcmin$. 2) The full BAT region is only partially covered by soft X-ray observations, and no X-ray counterpart is found. For those 9 sources, BASS 157 catalog has used old association held over from the previous catalog (see \cite{2025arXiv250604109L}), and their proposed counterparts do indeed fall into the region not covered by soft X-rays.
    \item The remaining 14 sources have different counterparts in the 157-month catalog compared to this work. The reasoning is similar to that of the previous point: different radii. 
\end{itemize}

\begin{figure}[ht]
    \centering
    \includegraphics[width=0.47\textwidth]{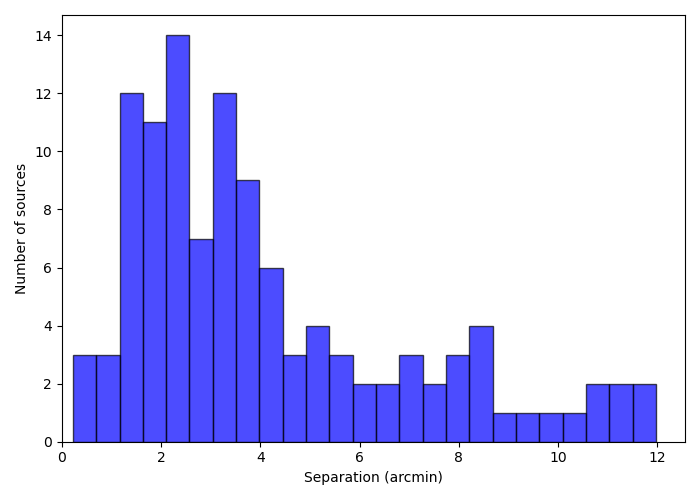}
    \caption{A histogram of separation between the BAT counterparts in this work and the BAT coordinates of \textit{Swift}-BAT 157 month catalog sources. $78\%$ (89 out of 114) of the sources are present within $6\arcmin$ from the \textit{Swift}-BAT 157 month catalog sources. }
    \label{fig:Separation}
\end{figure}

Figure \ref{fig:Separation} represents a histogram of the separation between the counterparts of BAT sources presented in this work and 157-month BAT sources. 89 out of these 114 sources are present within $6 \arcmin$ of the \textit{Swift}-BAT 157--month catalog sources.

We note that the source detection methods employed by the BASS collaboration catalogs and the Palermo BAT catalogs are different, and that error estimates of the BAT data are non-trivial. This is likely the origin of some of the discrepancies in source detections between the different
catalogs, particularly in this low signal-to-noise regime. As such, it is not unexpected that not all the counterparts found in this work have an associated 157-month catalog BAT detection.

\subsection{Positional accuracy}

\begin{figure}[ht]
    \centering
\includegraphics[width=0.47\textwidth]{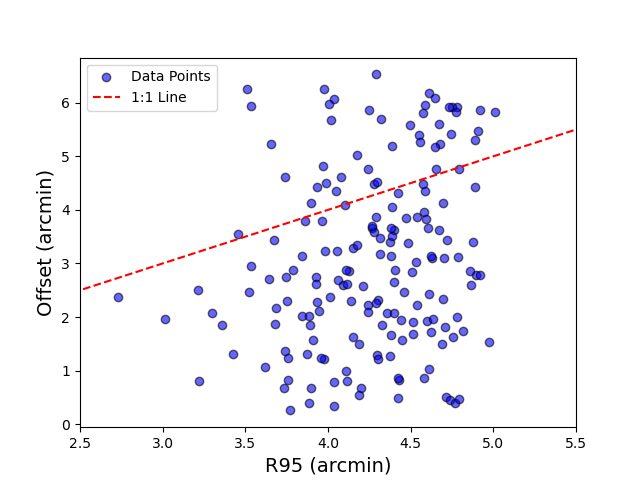}
    \caption{A plot showing distribution of BAT sources with BAT offset and BAT $R_{95}$. The angular separation between the counterpart position and the fitted BAT position is used to determine a measured BAT offset for each source. The majority of the sources lie below the identity line (1:1) indicating that most of the counterparts are well within the BAT $R_{95}$.}
    \label{fig:offsetR95}
\end{figure}

 Figure \ref{fig:offsetR95} presents a plot illustrating the distribution of BAT sources as a function of the BAT offset (i.e. the separation between the BAT source centroid and the position of the counterpart) and the BAT $R_{95}$. Among the 251 source counterparts, only 197 (78.5\%) are located within their respective $R_{95}$ regions, while 48 (19.1\%) fall between $R_{95}$ and 6\arcmin. Additionally, 6 sources ($\sim2.4$\% of the sample) are situated near the boundary of the {6\arcmin} circle, where the centroid of the soft X-ray detection lies outside the circle, but the emission extends into it. The median offset between the BAT source coordinates and the counterpart position is 2.85\arcmin, while the median radius of BAT $R_{95}$ is 4.30\arcmin. While the majority of the sources are positioned below the 1:1 identity line, our comparison between the $R_{95}$ and the 6' radius \citep[which was used for counterpart identification in previous iterations of the PBC,][]{2010A&A...510A..47S} demonstrates the existence of a small systematic uncertainty not accounted in the $R_{95}$ values.
 Indeed, if $R_{95}$ properly accounted for all sources of error, 95\% of the counterparts would fall within it. As such, we recommend using the 6' radius when searching for counterparts within this catalog, as well as previous, and future iterations of the PBC catalog.

\subsection{Cross-match with the 4FGL catalog}\label{4fglcrossmatch}

\begin{table*}
\centering
\caption{ Summary of the sources in our sample with a counterpart in the 4FGL-DR4 catalog. \label{4fgl}}
\begin{tabular}{ccccc}
\hline\hline
Name & Counterpart & Possible 4FGL Association & Counterpart to 4FGL source & Type \\ 
\hline\hline
J0428.8--6705 & SWIFT J042749.8-670434 & 4FGL J0427.8-6704 & 1SXPS J042749.2-670434 & LMB \\
J0640.3--1256 & OKM2018 SWIFT J0640.3-1286 & 4FGL J0640.0-1253 & TXS 0637-128 & bll  \\
J0710.8-3855 & ICRF J071043.6-385037 & 4FGL J0710.8-3851 & AT20G J071043-385037 & fsrq\\
J1159.5+2913& Ton 599 & 4FGL J1159.5+2914 & Ton 599 & fsrq \\
J1419.6-6052 & PSR J1420-6048 & 4FGL J1420.0-6048 & PSR J1420-6048 & PSR\\
J1440.9--3844 & 2MASS J14403782-3846549 & 4FGL J1440.6-3846 & 1RXS J144037.4-384658 & bll \\
J1452.6--1317 & PMN J1452-1319 & 4FGL J1453.0-1318 & TXS 1450-131 & bcu \\
J1736.3--4443 & NGC 6388 & 4FGL J1736.2-4443 & NGC 6388 & glc \\
J1737.3+0235 & MITG J1737+0236 & 4FGL J1738.0+0236 & PKS 1735+026 & bcu \\
J1813.2--1248 & PSR J1813-1246 & 4FGL J1813.4-1246 & PSR J1813-1246 & PSR \\
J1816.6--3912 & 2MASS J18163594-3912464 & 4FGL J1816.1-3908 & unassociated & unknown \\
J1817.3-1642 & SWIFT J181723.1-164300 & 4FGL J1816.2-1654c & SNR G014.1-00.1 & spp \\
J2221.9+5952 & SWIFT J2221.6+5952 & 4FGL J2221.9+5955 & unassociated & unknown \\
 \hline
 \end{tabular}
%vspace{-0.7cm}
\tablefoot{The table contains 4FGL associations to our BAT-detected sources, and lists their counterpart and source type as they appear in the 4FGL-DR4. The source types are as follows; LMB: Low-mass X-ray binary. bll: BL Lac type of blazar. fsrq: FSRQ type of blazar. PSR: pulsar, identified by pulsations. bcu: active galaxy of uncertain type. glc: globular cluster. spp: special case - potential association with SNR or PWN. We note that, although the counterpart names listed for the BAT and the 4FGL sources may differ, they are associated to the same source in all cases but one: J1817.3-1642. See Sect. \ref{4fglcrossmatch} for details.}
\end{table*}

Table \ref{Counterpart types} lists two sources as ``GammaS''. However, these are only classified as such by \textit{INTEGRAL} \citep{2007ApJS..170..175B} in NED/SIMBAD. Due to the  bandpass of \textit{INTEGRAL} (energy range 15  keV to 10 MeV, which fully includes the BAT energy range), these are more likely hard X-ray sources rather than genuine gamma-ray emitters. One of them (IGR J13045-5630) is only detected by BAT, Integral, and \textit{Swift}-XRT and lays outside of the Galactic plane, making it likely to be an AGN of unknown type (including, potentially, a blazar). The other, SWIFT J042749.8-670434, is detected at both UV and IR wavelengths, but since it lays within the plane, its type is uncertain. 

To obtain genuine gamma-ray counterparts to the newly-detected sources, we cross-matched the catalog presented here against the \textit{Fermi} 4FGL--DR4 catalog \citep{2022ApJS..260...53A,2023arXiv230712546B}. To do so, we use our BAT counterpart positions and a 30'' uncertainty radius, against the 4FGL-DR4 catalog, considering the semi-major axis of their 95\% uncertainty region. This reveals 13 sources with potential gamma-ray counterparts, as shown in Table \ref{4fgl}. Among these, there are two known BL Lacs (bll), two sources classified as flat-spectrum radio quasars (fsrq), two ``active galaxies of uncertain type'' (bcu, which are most likely to be a blazar of uncertain type), two pulsars (PSR), a globular cluster (glc), a low-mass X-ray binary (LMB), a possible association to a supernova remnant or a pulsar wind nebula (spp),and two unassociated 4FGL sources (4FGL J1816.1-3908 and 4FGL J2221.9+5955). 

Table \ref{4fgl} lists both the BAT counterparts suggested in this work, and those reported in the 4FGL. Despite the different names used, 11/13 of them are known associations of the same source, and the separation between our XRT--detected, X-ray counterpart and the counterpart reported in the 4FGL catalog is always $\lesssim$2$^{\prime\prime}$. The first exception is J1737.3+0235 (or 4FGL J1738.0+0236), for which we propose the radio source MITG J1737+0236 as the most likely counterpart, and the 4FGL lists PKS 1735+026. The two sources are separated by 6$^{\prime\prime}$, and are not associated with each other in SIMBAD. However, the Parkes Southern Radio Source Catalog \citep{Wright1990} lists the detection error of MITG J1737+0236 to be as high as 20$^{\prime\prime}$, making it likely that even for this object, the two counterparts are in fact the same source.

The second exception is J1817.3-1642, for which we list an X-ray source of unknown origin (SWIFT J181723.1-164300) as the counterpart of the BAT source, while the 4FGL lists SNR G014.1-00.1 (a supernova remnant) as the most likely association, and PMN J1816-1649 (a radio source) as a less likely (but still possible) association. Neither of these two sources fall within the BAT 6$^{\prime}$ uncertainty region, although the centroid of the SNR lays about $\sim3$$^{\prime}$ away. The SNR is not detected in the \textit{Swift}-XRT range, and the radio source, which lies more than 10$^{\prime}$ away, is not covered by the observation. All three sources, however, fall within the $R_{95}$ of the \textit{Fermi}-LAT source. Due to its intense brightness and hardness in X-rays (average $0.3-10$~keV flux $\sim3\times10^{-10}$~erg~s$^{-1}$ and $\Gamma\sim1.5$), as well as its BAT association, we suggest that SWIFT J181723.1-164300 is also a candidate counterpart to 4FGL J1816.2-1654c.

The coincidence of at least 10/11 of the BAT and 4FGL counterparts means that the spatial coincidence between a BAT and a 4FGL detection is unlikely coincidental. This implies that, for the last two sources in Table \ref{4fgl} (which lack an association in the 4FGL catalog), the listed BAT counterparts found in this work are likely to be associated with the Fermi source as well. They both lay in the Galactic plane, making it impossible to provide a likely source type without further multiwavelength follow up observations. 

\subsection{Optical follow-up of unclassified counterparts}
 
As mentioned in previous sections, a number of the associated counterparts are optically classified as galaxies, or infrared sources. This is due to the lack of an available optical spectrum, which would most likely confirm their AGN nature, as well as provide a redshift measurement for them.

In this section, we present the results of a short optical spectroscopic campaign, aimed at classifying 9 of these sources. We intend to maintain an updated list of any new optical follow-ups on our website \footnote{https://science.clemson.edu/ctagn/optical-follow-up-of-pbc-150-month-sources/}.

\begin{figure}[ht]
    \centering
\includegraphics[width=0.49\textwidth]{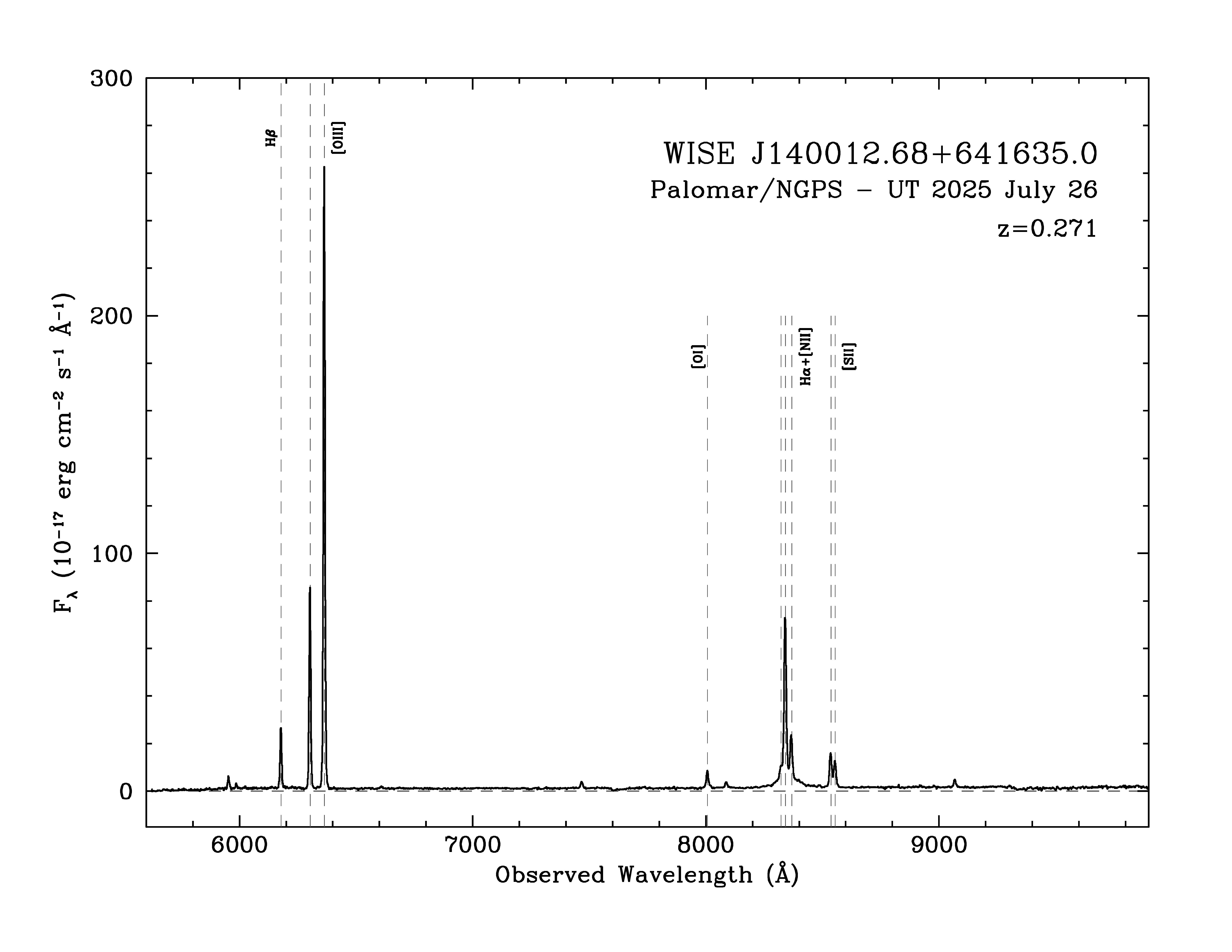}
    \caption{NGPS spectrum of WISEA J140012.68+641635.0, showing the narrow lines typical of a Seyfert 2 AGN.}
    \label{fig:Palomar}
\end{figure}

Five of the sources were observed with the Southern Astrophysical Research (SOAR) telescope, using the Goodman High Throughput Spectrograph \cite{Clemens2004}. The other four were observed using Next Generation Palomar Spectrograph (NGPS) at the Hale Telescope. The results of the follow-up can be found in Table \ref{optical_follow_up}, where we list the BAT name, the counterpart name, and its optical type and redshift according to the spectra obtained. Figure \ref{fig:Palomar} shows one of the NGPS spectra for reference, while the others can be found on the aforementioned website.

We note that 8 out of 9 sources have a redshift above the median of all BAT catalogs (including the sample presented here), and 5 of them have redshift values among the highest of the BAT catalog for non-beamed AGN ($z>0.2$). For context, the highest-$z$ Sy2 in the current Swift-BAT 157-month of \citet{2025arXiv250604109L} has $z=0.5975$, while WISEA J213603.27-162015.3 presented here has $z=0.611$. Only 46 non-beamed AGN (only 8 of which are Sy2) have $z>0.2$, and only 19 are at $z>0.3$.

This highlights the importance of optical follow-ups of fainter BAT sources, like the ones presented in this work.

{\bf
\begin{table*}
\centering
\caption{ Summary of the originally unclassified sources in our sample that were followed-up with optical spectroscopy. \label{optical_follow_up}}
\begin{tabular}{ccccc}
\hline\hline
Name & Counterpart & Optical Type  & Redshift & Instrument\\ 
\hline\hline
J0636.5+5911 & GALEXASC J063634.15+591319.6  & Sy2 & 0.205 & NGPS \\
J1125.5-3905 & WISEA J112551.03-390317.8 & Sy2 & 0.0298 & Goodman \\
J1400.5+6419 & WISEA J140012.68+641635.0 & Sy2 & 0.271 & NGPS\\
J1419.7-0427 & WISEA J141945.76-043016.4 & Sy1 & 0.3090 & Goodman\\
J1432.4+4141 & SDSS J143214.19+414004.8 & Sy1.9 & 0.216 & NGPS\\
J1616.3-1031 & LEDA 3082653 & Sy2 & 0.0653 & Goodman\\
J1626.7-3307 & WISEA J162642.27-330522.1 & Sy2 & 0.1090 & Goodman\\
J1917.2-2836 & WISEA J191649.60-283201.5 & Sy1 & 0.1739 & Goodman\\
J2135.9-1617 & WISEA J213603.27-162015.3 & Sy2 & 0.611 & NGPS \\
 \hline
 \end{tabular}
%vspace{-0.7cm}
\tablefoot{The table contains optical classification and redshift as derived from the optical spectra taken either using SOAR-Goodman, or Palomar.}
\end{table*}
}

\section{Conclusions}
 The \textit{Swift}-BAT 150--month catalog has a total of 2,339 detected sources, of which 344 are new hard X-ray detected sources without a low-energy counterpart.
  
 This work benefits from an extensive analysis of soft (2-10\,keV) X-ray data and cross-identification process to provide the classification for 250 out of these new 344 objects, and link them to optical/soft X-ray/infra-red counterparts using data from \textit{Chandra}, \textit{XMM-Newton}, \textit{Swift}-XRT, and \textit{eROSITA}.

This catalog includes beamed AGNs (blazars/FSRQ), non beamed AGNs (Galaxies),
cataclysmic variable stars (CVs), pulsars and young stellar objects which are important references for many scientific studies. Follow-up observations of BAT sources have always been crucial for the scientific community. \cite{2015MNRAS.454.3622B} showed correlations between X-ray continuum emission
and optical narrow emission lines based on the optical spectroscopic follow-up project \citep[the BAT AGN Spectroscopic Survey;][]{2017ApJ...850...74K}. The relationship between optical narrow-line emission line ratios and the Eddington accretion rate was investigated by \cite{2017MNRAS.464.1466O}. \cite{2017MNRAS.467..540L} also explored near-Infrared (NIR; 0.8--2.4 $\mu$m) spectroscopic properties of 102 Swift-BAT selected AGNs. Several works have investigated BAT sources to find and characterize obscured AGN \citep[e.g.,][]{2019ApJ...872....8M,2019ApJ...871..182Z,2019ApJ...870...60Z,2021ApJ...922..252T,2021A&A...650A..57Z,2022ApJ...940..148S,2023A&A...676A.103S, 2025ApJ...979..130C}.

Future studies using this catalog, in combination with X-ray surveys (e.g., \textit{eROSITA}, \textit{Chandra}) and multi-wavelength facilities (e.g., \textit{JWST}, \textit{Rubin}) will further refine our understanding of AGN demographics and their role in cosmic structure formation. This catalog thus serves as an important resource for advancing our knowledge of the hard X-ray sky and the physics of extreme astrophysical environments.

\section{Acknowledgments}
The team acknowledge funding from NASA under contract 80NSSC22K1469.
This research has made use of the SIMBAD database, operated at CDS, Strasbourg, France. 
This research has made use of the NASA/IPAC Extragalactic Database (NED),
which is operated by the Jet Propulsion Laboratory, California Institute of Technology,
under contract with the National Aeronautics and Space Administration. This research has made use of data and/or software provided by the High Energy Astrophysics Science Archive Research Center (HEASARC), which is a service of the Astrophysics Science Division at NASA/GSFC.

This work is based on observations obtained with \textit{XMM-Newton}, an ESA science mission with instruments and contributions directly funded by ESA Member States and NASA. 

This research has made use of data obtained from the \textit{Chandra} Data Archive provided by the Chandra X-ray Center (CXC).

This work is based on data from eROSITA, the soft X-ray instrument aboard SRG, a joint Russian-German science mission supported by the Russian Space Agency (Roskosmos), in the interests of the Russian Academy of Sciences represented by its Space Research Institute (IKI), and the Deutsches Zentrum für Luft- und Raumfahrt (DLR). The SRG spacecraft was built by Lavochkin Association (NPOL) and its subcontractors, and is operated by NPOL with support from the Max Planck Institute for Extraterrestrial Physics (MPE). The development and construction of the eROSITA X-ray instrument was led by MPE, with contributions from the Dr. Karl Remeis Observatory Bamberg \& ECAP (FAU Erlangen-Nuernberg), the University of Hamburg Observatory, the Leibniz Institute for Astrophysics Potsdam (AIP), and the Institute for Astronomy and Astrophysics of the University of Tübingen, with the support of DLR and the Max Planck Society. The Argelander Institute for Astronomy of the University of Bonn and the Ludwig Maximilians Universität Munich also participated in the science preparation for eROSITA.

 This work is based on observations obtained at the Southern Astrophysical Research (SOAR) telescope, which is a joint project of the Minist\'{e}rio da Ci\^{e}ncia, Tecnologia e Inova\c{c}\~{o}es (MCTI/LNA) do Brasil, the US National Science Foundation’s NOIRLab, the University of North Carolina at Chapel Hill (UNC), and Michigan State University (MSU).

\bibliographystyle{aa}
\bibliography{sample631}

\end{document}